# Is There Any Social Principle for LLM-Based Agents?

Jitao Bai [1, a] *
[1] Tianjin University
Tianjin, China
* Corresponding author: [a] e-mail: jitaobai_123@tju.edu.cn

Simiao Zhang [2, b] *
[2] East China Normal University
Shanghai, China
* Corresponding author: [b] e-mail: conver334@gmail.com

Zhonghao Chen [3, c]
[3] Tianjin University
Tianjin, China
[c] e-mail: diogeneschen@tju.edu.cn

**Focus on Large Language Model based agents should involve more than "human-centered" alignment or application. We argue that more attention should be paid to the agent itself and discuss the potential of establishing tailored social sciences for agents.**

Large Language Models (LLMs) arose in recent years have lit the people's enthusiasm on artificial intelligence (AI). Profiting from the pretraining on tremendous amounts of text data, LLMs can behave more like humans than earlier AI, and thereby can execute more complex tasks that can hardly be done before. A common way for LLMs to execute complex tasks is to instantiate them to be task-specific agents and form a community, in which they interact with each other, just like the collaboration in humans. But that has brought about an issue in such LLM-based agent communities: is there any social principle for those human-like agents?

Recent studies mainly follow two mainstreams, i.e., how to further make LLM-based agents to behave more like humans, and how to use them to simulate the human society to speed the discoveries in social sciences. However, these "human-centered" alignment or application may have limited our understanding on more essential perspectives of LLM-based agents. Motivated by the establishment of social sciences for humans, we discussed the potential of tailored social sciences for agents covering both the microscopic individual and the macroscopic group behavior features, which is expected to provide a deeper insight on how the agents behave and how to deploy them to construct more efficient communities.

As a note for clarity, some remarkable facts about agent individual and community are briefly discussed respectively in Section 1 and 2. These discussions will serve as a preliminary to readers, and most importantly, the inspiration and starting point for our viewpoints in Section 3.

## I. AGENT INDIVIDUAL AND HUMAN BEING: SAME OR DIFFERENT?

LLM-based agents are instantiated from LLMs with some suitable prompts and are usually equipped with the ability of planning, memory, as well as tool use to form complete systems, in which the LLMs play a role as the brain for human. In this case, the behavior of LLM-based agents is largely governed by LLMs. LLMs like ChatGPT have left a deep impression on us due to the capability in generating human-like text and some level of cognition like people. That is achieved through the vast training on human data. Some strategies in imitation of human behavior also promote the appearance of such performance, like self-reflection [1]. In current LLMs, similar behavior to human in the perspectives of heuristics, biases, and other decision effects can be observed [2]. And with the update on model generation, LLMs like GPT-4 has shown more complex behavior that is generally found only in human, for example, understanding and inducing false beliefs in other LLMs [3]. Such similarity in behavior has risen more in-depth investigation on machine psychology. It is revealed that just like human, LLMs also have personality, and different personality in LLMs can be shaped in a controllable manner [4]. They are able to grasp various emotions and show accordingly some extent of empathy [5]. When provided with different incentives, LLMs will show different levels of trust to the interlocutors [6].

The findings in the behavior and psychology of LLMs have emerged the thinking on the connection between the two. One study showed that the harmful behaviors of LLMs can be corrected through psychotherapy [7], offering the preliminary evidence that intervening the consciousness in LLMs can lead to a different behavioral response. In this sense, differences in cognition will affect the way the LLMs act. Actually, differences do exist in the cognition of LLMs and human beings. In human society, personal values appear in a great diversity and can be strong and conflicting. While in LLMs, the values tend to be neutral, especially for larger models [8], mainly due to the pretraining on more comprehensive text. Such differences may bring different behavior patterns in agent individual, and further, in group level.

## II. AGENT COMMUNITY TARGETED AT HUMAN SOCIETY: HOW IT ACTS?

Despite the nonnegligible differences, LLMs will still reframe the methodology of social science research for their striking similarity to human beings, and LLM-based agents are expected to supplant human participants for data collection, enhance policy analysis considering different theoretical or ideological views, and serve as controlled experimental partner in social interaction research [9]. Currently, LLM-based agents are usually involved as communities in experiments concerning game theory, such as the iterated Prisoner's Dilemma or other economic scenarios like ultimatum game, dictator game, and public goods game [10], in which multiple agents are organized in a way of either cooperation or competition. Others may also use agent communities to simulate the human social networks. In these simulations, phenomena in human society like the propagation of information, attitudes, and emotions [11] were captured. But in addition to what the agents are expected to do, some emergent social behavior [12] can also be observed. That suggests the agents have some autonomy in the simulated networks and may behave differently from humans in a group level.

## III. WHAT ABOUT AGENT SOCIETY ITSELF?

The differences in the behavior and psychology of LLM-based agents and human beings have triggered the research on AI alignment, i.e., making the LLMs behave and think in a way closer to human. Then with the increasing similarity, the agents were used to simulate the human society. But when we meet with a different social behavior in the agent-modelled society, we will subconsciously compare it with what humans do and again set our goal to make the agents behave more like us in hope that human society could be simulated more precisely. Of course making the agents behave just as we humans do seems to be a growing consensus among the community, should their own features be neglected?

In human society, the psychology and values have determined the way one behaves, and the aggregation of individuals' behavior forms the macro behavior of the society. That macro behavior as well as the constituent individual features are abstracted and finally lead to what we call social sciences. Similarly, from the behavior of single agent, a set of "social sciences" for agent community may also be derived. Since there exist inherent differences in the way agents and human act, the "social sciences" for agent society may also be different from that for human society. As illustrated in Figure 1, where the symbol $S_h$ represents the social sciences for human and $S_a$ the social sciences for agents, $S_a$ gradually gets closer to $S_h$ due to the increasing similarity in agents and human beings. But owing to the inherent differences between agents and human, at least in the current generation of LLMs, $S_a$ and $S_h$ cannot be completely the same. It can be predicted that with the advances in AI alignment, $S_a$ will finally converge to $S_h$, as indicated by curve I. At that moment, the social principles for human will be fully applicable for agents, and the agents will also be able to precisely simulate the human society. There is also another possibility that $S_a$ may go as the curve II. This can happen when $S_a$ has reached a rather high similarity to $S_h$, that is, the agents are smart enough to evolve themselves so that social principles different from human society can be created. Certainly, agent communities in such situation may still evolve towards higher similarity to human society, and curve II is merely schematic for one possible variation as no one knows exactly how on earth agent communities would evolve. But one thing is for sure that the social principles for agents and human are less likely to be completely different, and at least they would share something in common as after all the LLMs used for agents are pretrained on human texts. A premise is also necessary for the case illustrated by curve II: there is no or little technical intervention from human exerted. In other words, we can either choose to regulate the agents to behave just like us humans (curve I) or allow them to act more freely (curve II), and the evolution of agent social sciences is largely dominated by technology policies. That has revealed another opportunity for the interdisciplinary research on both human and agent social sciences. Existing work mainly focuses on the technologies that can make $S_a$ closer to $S_h$ or using $S_a$ to represent $S_h$ to get more information about $S_h$. We have obtained much knowledge on $S_h$, but unfortunately, we still have no idea on what $S_a$ is, even though some facts about $S_a$ can be vaguely inferred from limited experimental evidence that $S_a$ may be scale- and task-guided [13].

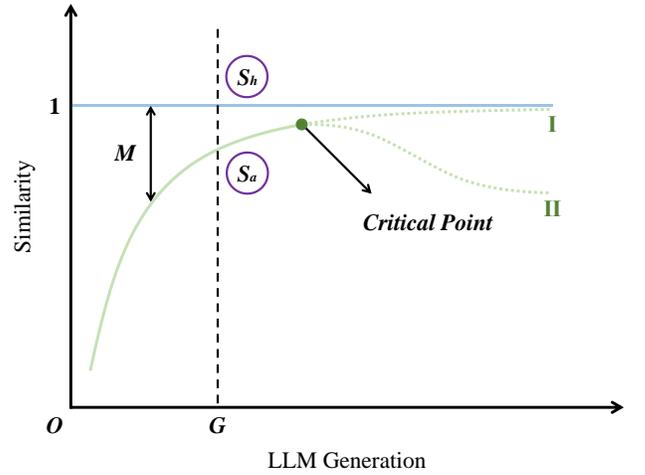

Figure 1. Evolution of social sciences for human ($S_h$) and agents ($S_a$). Similarity is established with the human social sciences serving as the baseline.

Similar to the common methodology in human social sciences, social sciences for LLM-based agents can be investigated both theoretically and empirically. Based on the behavioral and psychological features of individual agent, theories on agent society can be established with some reasonable hypotheses, just like the development of economics on the assumption that all the consumers are rational. These theories can be empirically tested and calibrated with the society constructed by numerous agents, which will finally lead to the set-up of agent social sciences. With the tailored social theory for LLM-based agents, we can have a deeper understanding on how and why they behave, and more importantly, give us a guidance on how to organize the agents to efficiently execute different tasks. Additionally, comparative study on human and agent social sciences is also encouraged. For example, metrics are needed to quantitatively measure the similarity between human and agent social sciences (variable $M$ in Figure 1), and the critical point for the possible transformation of the evolution mode of agent social sciences (highlighted in Figure 1) can be explored. Since human may also be involved in the same network with LLM-based agents to form a hybrid community, new theories may also be expected for hybrid human-agent social sciences. It should be clarified that the concept of agent social sciences discussed herein is never simply some specific agent behavior. Instead, it must be highly abstracted and generalized features or principles, forming a complete theoretical system. Here we give some basic criteria for the upcoming discipline of agent social sciences:

**Object**: Agent social sciences should be agent-centered, which requires agents to be the core object of the study in order to differentiate from human social sciences. Since the word "agent" is defined after human or animal in its philosophy origin as those with individual autonomy [14], we tend to rename it as "artificial intelligence entity (AIE)" to spin it off from the manifestation of human vassal.

**Scope**: Agent social sciences focus on both the individual (micro) and group (macro) behavior of LLM-based agents, their interaction with the environment, and the caused changes in the environment. Psychology of LLM-based agents is also included as an impartible connection with behavior. Human participation is allowed in the agent society investigated, just as the hybrid community of humans and agents mentioned before. In addition, the impact of human intervention like different technology policies on agent society as well as the comparative study on human and agent social sciences are also the domain or extension of agent social sciences.

**Paradigm**: Agent social sciences are a complete knowledge system with highly abstracted and generalized theories supported by empirical evidence. Simple observation can only serve as a preliminary for agent social sciences and the discipline should go further than that.

**Methodology**: Agent social sciences should be empirically-based. Though theoretical study is also an essential branch, the proposed theory must be tested empirically as previously discussed. Interdisciplinary methods may also be employed in the investigation, such as mathematical modeling.

**Ethics**: Study on agent social sciences mustn't do harm to human, neither physically nor mentally. Ethical review is necessary for projects that may threaten human ethical norms. It should also be ensured that the agents will never get out of control even when they have advanced to higher level of intelligence.

Social sciences for LLM-based agents have manifested a broad space for exploration. But till now, there are still the following challenges:

(1) Technical issues on LLM itself. As is known by the community, current LLMs still have some bugs. For instance, hallucination may arise in LLM-based agents [14], and the performance of LLMs on reasoning is still generally poor [15]. Moreover, some LLMs seem to have trouble in interaction, as they may forget the past conversations and may not distinguish important information from the irrelevant [16]. These cases can never be regarded as the nature of LLMs in agent social sciences since they can do nothing but disturbing the normal activities of agents.

(2) Level of prompts. Prompts are important in instantiating the LLMs to be particular agents. Insufficient prompts are not enough for agents to be the intended roles, while excessive prompts may cover up the essential features that should have been exhibited.

(3) Scale of agent community. Studies have found that the behavior of agents is related to their number in the community [13]. The number of agents involved in the reported experiments is generally small, which is unfavorable for the observation of group behavior and the scale effect must be considered.

(4) Distinguishing the essential attribute from what the LLMs have learnt. The behavior shown by the agents may not necessarily come from their essential attribute (i.e., the spontaneous behavior). Instead, the LLMs may have learnt how they should behave in the pretraining when faced with a similar task. Distinguishing such differences can be challenging.

(5) Evolution of agents. Self-evolution is a key factor that contributes to the transformation of agent society. Some emergent behavior has been reported in existing literature (e.g., know what tool to use when met with an unseen task and how to make the tool [17]). However, it is hard to say whether these agents have really evolved as they just seem to "memorize" more, while the parameters concerning the neurons in LLMs remain almost unchanged. In this sense, evolution of agents may need to be realized in broader dimensions.

The five issues must be addressed to pave the way for the development of agent social sciences. Maybe in a near future, we will be lucky enough to witness the plots of science fiction to come to reality, in which the intelligent agents build a parallel society to our human's.

**REFERENCES**

[1] Shinn N, et al. Reflexion: Language Agents with Verbal Reinforcement Learning. Preprint at *arXiv* https://doi.org/10.48550/arXiv.2303.11366 (2023).

[2] Suri G, Slater LR, Ziaee A & Nguyen M. Do Large Language Models Show Decision Heuristics Similar to Humans? A Case Study Using GPT-3.5. Preprint at *arXiv* https://doi.org/10.48550/arXiv.2305.04400 (2023).

[3] Hagendorff T. Deception Abilities Emerged in Large Language Models. Preprint at *arXiv* https://doi.org/10.48550/arXiv.2307.16513 (2023).

[4] Jiang G, et al. Evaluating and Inducing Personality in Pre-trained Language Models. Preprint at *arXiv* https://doi.org/10.48550/arXiv.2206.07550 (2023).

[5] Pataranutaporn P, Liu R, Finn E & Maes P. Influencing human-AI interaction by priming beliefs about AI can increase perceived trustworthiness, empathy and effectiveness. *Nature Machine Intelligence* 5, 1076-1086 (2023).

[6] Johnson T & Obradovich N. Measuring an artificial intelligence agent's trust in humans using machine incentives. Preprint at *arXiv* https://doi.org/10.48550/arXiv.2212.13371 (2022).

[7] Lin B, Bouneffouf D, Cecchi G & Varshney KR. Towards Healthy AI: Large Language Models Need Therapists Too. Preprint at *arXiv* https://doi.org/10.48550/arXiv.2304.00416 (2023).

[8] Zhang Z, et al. Heterogeneous Value Evaluation for Large Language Models. Preprint at *arXiv* https://doi.org/10.48550/arXiv.2305.17147 (2023).

[9] Grossmann I, et al. AI and the transformation of social science research. *Science* 380, 1108-1109 (2023).

[10] Phelps S & Russell YI. Investigating Emergent Goal-Like Behaviour in Large Language Models Using Experimental Economics. Preprint at *arXiv* https://doi.org/10.48550/arXiv.2305.07970 (2023).

[11] Gao C, et al. $S^3$: Social-network Simulation System with Large Language Model-Empowered Agents. Preprint at *arXiv* https://doi.org/10.48550/arXiv.2307.14984 (2023).

[12] Park JS, et al. Generative Agents: Interactive Simulacra of Human Behavior. Preprint at *arXiv* https://doi.org/10.48550/arXiv.2304.03442 (2023).

[13] Zhuge M, et al. Mindstorms in Natural Language-Based Societies of Mind. Preprint at *arXiv* https://doi.org/10.48550/arXiv.2305.17066 (2023).

[14] Xi Z, et al. The Rise and Potential of Large Language Model Based Agents: A Survey. Preprint at *arXiv* https://doi.org/10.48550/arXiv.2309.07864 (2023).

[15] Biever C. ChatGPT broke the Turing test-the race is on for new ways to assess AI. *Nature* 619, 686-689 (2023).

[16] Broadbent E, Billinghurst M, Boardman SG & Doraiswamy PM. Enhancing social connectedness with companion robots using AI. *Science Robotics* 8, eadi6347 (2023).

[17] Wang G, et al. Voyager: An Open-Ended Embodied Agent with Large Language Models. Preprint at *arXiv* https://doi.org/10.48550/arXiv.2305.16291 (2023).